\documentclass[lettersize,journal]{IEEEtran}
\usepackage{amsmath,amsfonts}
\usepackage{amssymb}
\usepackage{bm}
\usepackage{algorithmic}
\usepackage{algorithm}
\usepackage{array}
\usepackage[caption=false,font=normalsize,labelfont=sf,textfont=sf]{subfig}
\usepackage{textcomp}
\usepackage{stfloats}
\usepackage{url}
\usepackage{verbatim}
\usepackage{graphicx}
\usepackage{cite}
\usepackage{graphicx}
\usepackage{longtable}
\usepackage{caption}
\usepackage{multirow}
\usepackage{array}
\newcolumntype{P}[1]{>{\centering\arraybackslash}p{#1}}
\usepackage[dvipsnames]{xcolor}

\definecolor{NavyBlue}{rgb}{0.0, 0.0, 0.5} 
\definecolor{RoyalBlue}{rgb}{0.25, 0.41, 0.88} 
\definecolor{ForestGreen}{rgb}{0.13, 0.55, 0.13} 
\definecolor{LimeGreen}{rgb}{0.2, 0.8, 0.2} 
\definecolor{Firebrick}{rgb}{0.7, 0.13, 0.13} 
\definecolor{Crimson}{rgb}{0.86, 0.08, 0.24} 
\definecolor{Salmon}{rgb}{0.98, 0.5, 0.45} 

\begin{document}

\title{Traffic Engineering in Large-scale Networks with Generalizable Graph Neural Networks}

\author{Fangtong Zhou*, Sihao Liu*, Xiaorui Liu, Ruozhou Yu, Guoliang Xue
\thanks{
Fangtong Zhou, Sihao Liu, Xiaorui Liu and Ruozhou Yu(\{fzhou, sliu78, xliu96, ryu5\}@ncsu.edu) are with the Department of Computer Science, North Carolina State University, Raleigh, NC, 27606, USA.
Guoliang Xue (xue@ncsu.edu) is with the School of Computing and Augmented Intelligence, Arizona State University, Tempe, AZ, 85287, USA.
The research of Fangtong Zhou, Sihao Liu, and Ruozhou Yu is supported in part by NSF grant 2045539 and 2433966. 
The research of Xiaorui Liu is supported by the NSF grant 2443182.
The information reported here does not reflect the position or the policy of the funding agency. This paper has been accepted for publication in IEEE/ACM Transactions on Networking.\\
*~The first two authors contributed equally to this paper.

}

}

\maketitle

\begin{abstract}

Traffic Engineering~(TE) in large-scale networks like cloud Wide Area Networks~(WANs) and Low Earth Orbit~(LEO) satellite constellations is a critical challenge.
Although learning-based approaches have been proposed to address the scalability of traditional TE algorithms, their practical application is often hindered by a lack of generalization, high training overhead, and a failure to respect link capacities. 
This paper proposes TELGEN, a novel TE algorithm that learns to solve TE problems efficiently in large-scale network scenarios, while achieving superior generalizability across diverse network conditions. TELGEN is based on the novel idea of transforming the problem of ``predicting the optimal TE solution'' into ``predicting the optimal TE algorithm'', which enables TELGEN to learn and efficiently approximate the end-to-end solving process of classical optimal TE algorithms.
The learned algorithm is agnostic to the exact underlying network topology or traffic patterns, and is able to very efficiently solve TE problems given arbitrary inputs and generalize well to unseen topologies and demands.
We train and evaluate TELGEN with random and real-world topologies, with networks of up to $5000$ nodes and $3.6 \times 10^{6}$ links in testing.
TELGEN shows less than $3\%$ optimality gap while ensuring feasibility in all testing scenarios, even when the test network has $2\text{-}20\times$ more nodes than the largest training network.
It also saves up to $84\%$ TE solving time than traditional interior-point method, and reduces up to $79.6\%$ training time per epoch than the state-of-the-art learning-based algorithm.
Our code is available at \url{https://github.com/Amihua/TELGEN}.
\end{abstract}

\section{Introduction}
\label{sec:intro}

\noindent
Traffic Engineering (TE) is becoming increasingly crucial amid the exponential growth in Internet traffic.
With the help of software-defined networking (SDN), TE sits at the center of network management, supporting billions of applications and users for achieving the best performance and experience.
Usually, TE is implemented by a central
controller that has a global view of the network and can make informed decisions about routing and traffic splitting to optimize traffic~\cite{te_survey}.

Crucial challenges for the central
controller to manage TE are diverse network conditions such as the traffic pattern and network topology changes, and scalability issues in large-scale networks.
Depending on the network scenario, network traffic patterns are governed by time-varying user activities, automated machine tasks, natural, social and political events, and so on.
The topology could also vary because of network congestion, outages, or unexpected failures, which can significantly affect network performance such as throughput and quality-of-service (QoS) that can further lead to higher operational costs, revenue loss or penalty, if not dealt with promptly.
Tackling this challenge requires periodically re-configuring network TE in the time scale of minutes to hours, which poses a significant overhead on the central controller~\cite{teal}.
%
%
%

The scale of modern computer networks exacerbates the challenge of TE.
Current Wide-Area Networks~(WANs), such as Cloud-WAN or Low Earth Orbit~(LEO) satellite backbone network, typically consist of hundreds to thousands of nodes and tens of thousands to millions of links.
These networks present prohibitively large TE problem sizes that existing optimization-based algorithms cannot catch up with.
State-of-the-art Linear Program (LP) solvers, widely used to solve TE problems, can lead to up to cubic complexity in the number of links and traffic flows, which is prohibitive for modern network sizes.

Some decomposition-based algorithms have been proposed~\cite{teal2, teal46}, but can only utilize limited CPU parallelism to achieve up to moderate speed-up with no asymptotic improvement~\cite{teal}.
With the emergence of large-scale and dynamic networks\cite{10.1145/3544216.3544265, zhang2021gemini}, classical TE faces fundamental challenges in terms of scalability and responsiveness. While traditional WAN TE cycles could tolerate latencies of several minutes, modern environments like large-scale data centers and Low-Earth Orbit satellite constellations now demand real-time responses on the order of seconds \cite{xu2025decouple}. For instance, in LEO networks, the topology changes so rapidly that over 50\% of network paths can become obsolete within 150 seconds \cite{wu2025sate}. State-of-the-art LP solvers, however, can require tens of seconds for computation alone, a timeframe that renders them ineffective for such dynamic scenarios \cite{wu2025sate}. This creates a critical need for a new class of TE solutions that can deliver near-optimal traffic allocations at millisecond latencies, thereby making real-time control loops feasible.


%
%
Recently, the power of machine learning (ML) has been explored to speed-up TE in large-scale networks.
By using ML algorithms such as graph neural network (GNN) or reinforcement learning (RL), networks can automatically learn and adapt to traffic patterns, predict congestion, and optimize routing in real-time. 
This results in improved network performance, reduced latency, and better resource utilization~\cite{rl_0},~\cite{rl_1},~\cite{rl_2},~\cite{rl_3}, making ML a promising tool for modern TE.
Despite the early success, however, the current learning-based TE algorithms have several inherit limitations, which significantly hinder their practical application in real-world TE scenarios.
First, existing algorithms typically involve training with various traffic demands on the same network topology as testing or inference, with at best minor topology changes such as individual link or node failures~\cite{teal, DOTE}.
In~\cite{trans_wan}, up to $20\%$ of the topology changes are allowed by the form of capacity variation considered as inactive links or subsets of nodes from one topology.
This leads to inferior generalizability of the trained models, making them perform poorly on unseen topologies (such as when deployed in a new network) or in face of significant topology changes (such as regional failures); they also do not generalize to traffic matrices that are substantially different from the training data which is not flexible when dealing with training and testing data from different ranges.
Second, many of these algorithms require extensive training time and computation power.
For instance, TEAL~\cite{teal} needs to be trained on the latest GPUs for more than a week for a large topology and HARP~\cite{trans_wan} has difficulty dealing with large topology with node number over $1000$ in the traffic matrix processing part.
This makes it very difficult to (re-)train models for different network scenarios and conditions.
Furthermore, as they adopt unsupervised or self‑supervised learning approaches, existing algorithms do not always respect TE constraints, for example link capacities, and must be combined with expensive classical algorithms such as Alternating Direction Method of Multipliers~(ADMM) to ensure feasibility of the predicted TE solutions.
To address the above challenges, we propose \textbf{TELGEN}, a novel end-to-end learning-based TE algorithm for large-scale networks.
TELGEN is designed with the goal of enabling generalizability to diverse network topologies, sizes, traffic patterns and conditions without or with minimal re-training.
To achieve both generalizability and constraint satisfaction, TELGEN employs a flexible GNN architecture that is agnostic to the topology or traffic demands, and is trained with \emph{(step-by-step) supervised learning} to mimic the most powerful classical TE algorithms with high efficiency and optimality.
Specifically, by casting TE problems as LPs, TELGEN learns the detailed steps that a classical TE algorithm (such as the interior-point method or IPM) takes to solve smaller-scale TE problems till optimality, and then generalizes the learned algorithm to larger networks and problem sizes utilizing the invariance properties of GNNs.
By employing strong supervision from the classical algorithm, TELGEN reduces the training overhead in terms of the amount of data and computation needed and speeds-up training by several orders of magnitude.
To evaluate its performance, we train and evaluate TELGEN on a wide range of randomly generated and real-world network topologies.
Experiments show that TELGEN can achieve close-to-optimal TE (with $<3\%$ optimality gap
while ensuring feasibility), is up to $7$ times faster than a state-of-the-art IPM solver, and reduces the training time per epoch by up to $79.6\%$ over TEAL and HARP.
Further, TELGEN maintains negligible optimality gap even when the test networks are up to $2$-$20\times$ larger than the largest network in training, or when demand patterns vary largely between training and testing, demonstrating superior generalizability of TELGEN in face of diverse network scenarios and conditions.
Our main contributions are as follows:
\begin{itemize}
    \item 
    We propose a flexible GNN architecture for modeling and learning efficient TE algorithms that is agnostic to network topology, size, condition and traffic demands.
    \item 
    We propose a general supervised learning approach for efficiently learning close-to-optimal and generalizable GNN-based algorithms for solving large-scale TE problems, by mimicking classical TE algorithm on smaller instances for the training process and testing on larger, diverse networks.
    \item 
    We conduct extensive simulations on both randomly generated and real network topologies with up to $5000$ nodes, and demonstrate the superior optimality, training efficiency, inference time of TELGEN, compared to state-of-the-art solver and learning-based algorithms.
    Especially, the generalizability over network topology, size, condition as well as traffic demand of TELGEN has been justified, compared to state-of-the-art solver and learning-based algorithms.
\end{itemize}

In the following, Sec.~\ref{sec:rw} discusses related works. 
Sec.~\ref{sec:model} describes the system model and problem formulation. 
Sec.~\ref{sec:algo} describes the proposed GNN architecture and the training algorithm for solving large-scale TE problems.
Sec.~\ref{sec:eval} shows evaluation results. 
Sec.~\ref{sec:conclusions} concludes this paper.

\section{Background and Related Work}
\label{sec:rw}
\subsection{Traffic Engineering}
\noindent
The primary goal of TE is to direct traffic within a data network to meet traffic demands by optimizing a chosen performance metric~\cite{te_survey}.
State-of-the-art WAN TE solutions rely on formulating it as LPs or convex programs, which can be solved by commercial solvers using standard optimization algorithms (such as IPM, simplex method or primal-dual methods).
However, the swift expansion and addition of new data centers or the emergence of new networks such as large-scale LEO satellite networks have made the TE optimization decision making process exceedingly slow and impractical at larger scales. 
Nowadays, it takes commercial solvers several hours to allocate traffic on WANs comprising thousands of nodes~\cite{teal}.
As a result, WAN operators are urgently looking for ways to expedite TE optimization to match the rapid growth of their networks.
Methods haven been proposed to parallelize LP solvers, which nevertheless only achieve a coarse-level parallelization over CPUs and at best a moderate speed-up~\cite{teal2, teal46}.
Approximation algorithms such as Fleischer’s algorithm~\cite{fleischer_approximating} have also been a solution to speed up TE, which nevertheless compromises optimality and can lead to many iterations to converge.
\subsection{Machine Learning for Traffic Engineering}
\noindent
Methods have emerged to utilize ML to address the scalability issue in TE, with some promising initial results on significantly accelerating TE optimization~\cite{te_ml_0, te_ml_1, te_ml_2, te_ml_3}. 
By training on historical or randomly generated data, ML models can make accurate predictions that enhance the efficiency of network management. 
Early approaches began with using Convolutional Neural Networks (CNNs) to efficiently handle spatial dependencies in network data~\cite{te_ml_2}, focuses on predicting traffic matrices as inputs to traditional TE algorithms. 
Subsequent works have explored deep RL to solve various TE problems including TE automation, delay minimization, power-awareness and dynamic changes of traffic demands~\cite{rl_0, rl_1, rl_2, rl_3}.
While promising, simple (deep) RL algorithms are difficult to encode or learn detailed network information such as topology and capacity constraints, and can lead to unstable performance or frequent capacity violations in practice.

In TEAL~\cite{teal}, the authors combined multi-agent RL (MARL) with a GNN-based feature extractor to learn network-specific information (such as topology and flows), and applied classical ADMM to ensure feasibility.
The GNN-MARL model is trained with historical or randomly generated demands on the target network, and performs reasonably well \emph{on the same network} with different demands.
However, TEAL can only be applied to the exact same set of source-destination (SD) pairs on the same network as training, and have to be re-trained from scratch even when a new node is added or an existing node is removed, significantly limiting its generalizability.
It also suffers from significant training time and overhead due to the use of RL, and cannot satisfy link capacity constraints without substantial help from ADMM.

Compared to RL-based solution, another approach is to directly learn a mapping from the problem inputs (or historical demands) to the TE solution, such as in DOTE~\cite{DOTE} and FIGRET~\cite{FRIGHT}.
These algorithms directly learn and encode the network topology into the neural model parameters, and hence cannot generalize well on topologies beyond the training set.
A related recent paper presents a neural TE model, HARP, that can generalize to changing topologies~\cite{trans_wan} with a topology variation up to $20\%$ and the graph size up to less than 1000 nodes.
HARP considers continuous topology changes during TE execution, and does not to tackle the issue of TE solution generalizability to topologies significantly different from or larger than those in the training dataset.
As such, the approach would not be able to reduce training overhead by training on smaller networks and then applied on much larger ones, or instantly be applied when significant topology changes such as regional failures happen.

\noindent
\subsection{Graph Neural Network}
\noindent
GNNs~\cite{gnn_survery} have emerged as a powerful framework for machine learning on graph-structured data. 
Let a graph be represented as $G = (V, E)$, where $V$ is the node set and $E$ is the edge set of the graph.
The neighbor node set of a node $v$ is defined as $N(v) = \{u \in V |(v, u) \in E\}$.
Define the adjacency matrix $A^{n \times n}$ with $A_{ij} = 1$ is $e_{(i, j)} \in E$, and 0 otherwise, with $n = |V|$.
Let $h_v$ be the node attribute for $v \in V$.
A GNN model consists of a fixed number of GNN layers, denoted by $K$, and let a node's attribute after layer $k$ be $h_v^k$.
In each layer $k$, a message passing function denoted as $MSG^k(\cdot)$ is applied to pass and aggregate the node attributes of all neighbors of each node $v$, and an aggregation function denoted as $g^k(\cdot)$ is then applied to aggregate the aggregated neighbor messages with the node's own attribute $h_v^{k-1}$, outputting the updated node attribute $h_v^k$.
A GNN layer can thus be represented as:
\begin{equation}
\label{gnn_layer}
    h_v^k = g^k (h_v^{k-1}, MSG^k (\{h_u^{k-1} | u \in N(v)\}) ).
\end{equation}
To make a prediction, a readout function $r(\cdot)$ is further applied to node attributes after layer $K$ to predict the target node, link or graph-level labels.
In a GNN, the functions $MSG^k$, $g^k$ and $r$ are modeled by parameterized neural networks (such as multi-layer perceptrons or MLPs), whose parameters are trained with training data guided by a loss function.
GNNs are increasingly utilized both as direct solvers and to augment  existing solvers~\cite{co_gnn}.
In~\cite{tsp}, a learning-based method for solving the Travelling Salesman Problem has been proposed via deep graph convolutional networks.
An imitation learning framework of branch-and-bound has been proposed by~\cite{branch-and-bound} to parameterize the state of the branch-and-bound search tree for heterogeneous Mixed-Integer Linear Programming problems.
Likewise, \cite{co_nphard} tackled combinatorial problems that can be reduced to the Maximum Independent Set, Minimum Vertex Cover, Maximal Clique and Satisfiability using GNN models.
Follow-up research such as~\cite{co_gnn} studied solvers augmented by RL and observed that the GNN component is crucial for achieving competitive solution quality.
Despite their general popularity, GNNs are used in most existing works to augment other solvers, rather than as end-to-end solvers themselves.
%
%

%
%

\section{System Model and Problem Formulation}
\label{sec:model}


\noindent
The goal of TE is to allocate network resources among clients and optimize for network performance, such as maximizing throughput, minimizing latency, or minimizing maximum link utilization (MLU).
The widely used TE formulation involves splitting traffic demands between source-destination (SD) pairs along multiple paths.
This involves determining the best routes for data packets to travel and balancing the load across various network paths to avoid congestion and ensure efficient utilization of bandwidth. 
Our work focuses on traffic engineering in large-scale point-to-point networks, which is motivated by challenges observed in practical settings such as data centers and wide-area communication systems. In general, we believe our methodology can be applied to wireless scenarios and will study this in our future work.
%
Without loss of generality, we assume the objective of the network operator is to maximize the overall network throughput when meeting the demands of network SD pairs, subject to the link capacities.
\subsection{Network Model}

\noindent
\textbf{Network.}
The network is denoted as a directed graph $\mathcal{G} = (N, L, C)$, where $N$ is the set of nodes, $L$ is the set of links between nodes, and $C: L \rightarrow \mathbf{R^+}$ denotes the capacity of each link; $\mathbf{R}^+$ is the non-negative real number set.
There are in total $|N|$ nodes and $|L|$ links in $\mathcal{G}$.

\noindent
\textbf{Client demands.}
We consider a set of SD pairs, $(s_i, t_i, d_i) \in D$, where $s_i$ and $t_i$ are the source and destination nodes of SD pair $i$ respectively, and $d_i \in \mathbf{R}^+$ is the traffic demand; we use $i$ and $(s_i, t_i, d_i)$ interchangeably when referring to an SD pair.
We assume the demands are either given as input, or predicted by a traffic prediction model independent from the TE system.

\noindent
\textbf{Predefined network paths.}
The set of network paths for an SD pair $i$ is denoted by $P_i$.
Without loss of generality, we assume no two SD pairs have the same source and destination $(s_i, t_i)$, and define $P = \bigcup_i P_i$.
We write $n \in p$ and $l \in p$ to denote a node and a link that a path $p$ passes through, respectively.
Following most existing TE solutions~\cite{teal, DOTE, FRIGHT}, we assume the set of paths for each SD pair is given as input to the TE system.
It can, for instance, be calculated by Yen's $k$-shortest path algorithm~\cite{k-shortest} between each SD pair.
\subsection{TE Optimization Problem Formulation}
\noindent
\textbf{Traffic Splitting.}
The basic TE problem has decision variables,
defined as \emph{traffic splitting ratios} $\mathcal{R}$, which denote allocated fractions of each SD pair's demand to possible paths of the SD pair.
We define $\mathcal{R}(p)$ as the fraction of the demand $d_i$ that is allocated to path $p \in P_i$.
The total traffic split ratio of an SD pair cannot exceed $100\%$.
We thus represent the variables as a mapping $\mathcal{R}: P \rightarrow [0, 1]$, and enforce this constraint in the formulation as described below. 

\noindent
\textbf{Objective.}
The basic goal of TE optimization is to maximize network flow for all demands.
This lead to the objective function as: maximize $f_0(\mathcal{R}) \triangleq \sum_{i \in \textit{D}} d_i  \sum_{p \in P_i} \mathcal{R}(p) $.
We use this throughput maximization objective as the primary example in our description and evaluation, although our designed solution can be easily adopted to other linear objectives (such as minimizing MLU or routing cost), and possibly convex non-linear objectives as well (such as fairness).

\noindent
\textbf{Constraints.}
The problem has two sets of constraints.
First, the total allocated traffic for each SD pair cannot exceed its demand, and hence $\sum_{p \in P_i} \mathcal{R}(p) \leq 1$ for every $i \in D$.
Second, the traffic allocated on each link cannot exceed its capacity, and we have $\sum_{i \in D} d_i \sum_{p \in P_i: l \in p}  \mathcal{R}(p) \leq C(l)$ for $\forall l \in L$.

\vspace{0.02in}
\noindent
\textbf{TE Optimization.}
We formulate TE optimization as: 
\begin{equation}
\begin{aligned}
\label{TE}
\text{maximize} 
&\quad f_0(\mathcal{R}) \triangleq \sum_{i \in \textit{D}} d_i \sum_{p \in P_i} \mathcal{R}(p)\\
\textrm{subject to} 
&\quad \sum_{p \in P_i} \mathcal{R}(p) \leq 1, \quad \forall i \in \textit{D};\\
&\quad \sum_{i \in D} d_i     \!\!\!\sum_{p \in P_i: l \in p}\!\!\!\!\!  \mathcal{R}(p) \leq C(l), \quad \forall l \in L;\\
&\quad \mathcal{R}(p) \geq 0, \quad \forall p \in P.
\end{aligned}
\end{equation}

\noindent For simplicity, we define constraint functions $f^D_i(\mathcal{R}) \triangleq \sum_{p \in P_i} \mathcal{R}(p) - 1$, and $f^L_l(\mathcal{R}) \triangleq \sum_{p \in P_i: l \in p}  \mathcal{R}(p) - C(l)$.
We are interested in designing an algorithm that solves~\eqref{TE} to its optimality, ensures feasibility of the solution, and can do so efficiently on very large network scenarios (with thousands of nodes) even on those not observed in the training process.

\section{The TELGEN Learning-based TE Framework}
\label{sec:algo}

\noindent 
Diverse network conditions such network topology changes and dynamic demands complicate the task of efficiently TE. 
Additionally, scalability issues arise as networks grow in size, making it difficult to apply traditional TE methods that struggle to adapt and perform in real-time. 
These challenges underscore the need for innovative solutions that can dynamically respond to fluctuating network conditions and scale seamlessly with network growth. 
Therefore, we design our TE solution with a focus on addressing these two critical challenges—diverse network conditions and scalability. 
Our goal is to develop a robust framework that enhances TE, ensuring efficient and reliable data flow in increasingly complex network environments.
\subsection{Motivation and Overview}
\label{sec:motivation}

\noindent 
Compared to prior work, our first-class goal is to learn \emph{generalizable} neural models that can efficiently solve Program~\eqref{TE} on different network conditions, including different topologies, link capacities and demand distributions.
In principle, one can achieve these goals by developing models that can work with different sets of inputs, and then training these models with diverse training data to achieve generalizability.
Unfortunately, most ML models (\emph{e.g.} MLP, CNN or Transformer) do not have such flexibility due to their fixed input sizes or shapes, rendering solutions such as DOTE~\cite{DOTE} intrinsically non-generalizable.
Compared to other models, GNNs present a natural architecture that is invariant to permutation of input data, which means the output will not change when given such inputs, making them suitable to train generalizable models as demonstrated on various graph-alike data~\cite{gnn_generalization}.
This motivates us to propose a general TE framework using GNNs.

To apply GNNs for TE, a naive idea is to use a GNN to directly model the network structure.
This has been adopted in network performance modeling, e.g., utilizing GNNs to predict network queueing delays~\cite{Routenet-erlang}.
However, it is not trivial to extend this idea to modeling the correlations among routing paths, traffic allocation and link capacities, as each path may consists of multiple links; it is also unclear how to represent a global objective function such as the total throughput or the maximum link utilization.
Our solution thus develops a new GNN architecture, which not only represents the network topology, but also the correct correlations among paths, links and network-wide objectives in one graph.
This is achieved by transforming the LP formulation in Program~\eqref{TE} into a graph, thus preserving all the information in the TE formulation. 
To train a neural TE model, existing works rely on the TE objective and/or constraints as the only supervising signal for training, by defining them directly as the loss function or the reward function in RL.
However, this can lead to extremely long training epochs, and the trained model can usually reach a local optimum wich is not necessarily a global optimum. 
To address this issue, we draw inspiration from recent advances in ML on \emph{algorithmic alignment}~\cite{gnn_reason}, and develop GNN models that are directly \emph{aligned to} classical optimization algorithms (such as IPM). 
These models can then be trained directly with \textbf{strong supervision} from the intermediate steps generated by the classical algorithms, both achieving efficient training and ensuring convergence to the global optimal solutions that are output by these algorithms.
Since these classical algorithms are \emph{agnostic} to the network structure or input distributions---they are designed to solve any TE instance to optimality---the trained models will naturally learn such behaviors during training, thereby achieving inherit generalizability.

Next, we explain detailed design of our solution, and use IPM as the classical algorithm for supervision.
For completeness, we start with a brief introduction to the state-of-the-art IPM algorithm, and then describe our solution in details.

\subsection{IPM Preliminaries}

\noindent
IPM~\cite{ipm} is an efficient algorithm for optimally solving LPs and convex programs, and is widely adopted in commercial solvers~\cite{gurobi, 2020SciPy-NMeth, cplex}.
Consider a standard LP formulated as $\max_{\mathbf{x}} \{ \mathbf{c}^{\sf T} \mathbf{x} \,|\, \mathbf{A} \mathbf{x} \le \mathbf{b}, \mathbf{x} \ge 0\}$ with coefficient matrix $\mathbf{A}$ and constant vectors $\mathbf{b}, \mathbf{c}$.
A standard IPM~\cite{ipmgnn} is shown in Algorithm~\ref{algo:ipm}.
It starts from an initial strictly feasible solution $\mathbf{x}_0$, and then traverse the interior of the feasible region to find the optimal solution.
At each iteration, a search direction is determined, and a step size is calculated to ensure the new solution remains feasible while reducing a barrier function parameter $\mu$. 
The barrier parameter $\mu$ guides the solution towards the optimal point by penalizing infeasibility. 
The iterations continue until $\mu$ is sufficiently small, at which point the current solution $\mathbf{x}_k$ is considered approximately optimal.
\begin{algorithm}[t]
\caption{\mbox{IPM for Linear Programming}}
\begin{algorithmic}[1]
\label{algo:ipm}
    \REQUIRE Problem data $\mathbf{A}$, $\mathbf{b}$, $\mathbf{c}$; initial barrier parameter $\mu_0 \!>\! 0$; error tolerance $\epsilon \!>\! 0$; discount factor $\tau \!\in\! (0,1)$
    \STATE Choose initial feasible solution $\mathbf{x}_0$ such that $\mathbf{A}\mathbf{x}_0 \!\!<\!\! \mathbf{b}$ \\ and $\mathbf{x}_0 > 0$\;
    \STATE Set the initial barrier parameter $\mu \leftarrow \mu_0$, and $k \leftarrow 0$\;
    \WHILE{$\mu > \epsilon$}
        \STATE $k \leftarrow k + 1$\;
        \STATE Compute Newton direction $\Delta \mathbf{x}_k$ on barrier problem $\max_{\mathbf{x}} \{ \mathbf{c}^{\sf T} \mathbf{x} - \mu \left( \mathbf{1}^{\sf T} \log \left( \mathbf{b} - \mathbf{A} \mathbf{x} \right) + \mathbf{1}^{\sf T} \log (\mathbf{x}) \right) \}$\; \label{l:newton}
        \STATE Perform a line search to determine step size $\alpha_k$ \\ that ensures that $\mathbf{x}_{k} = \mathbf{x}_{k-1} + \alpha_k \Delta \mathbf{x}_k$ \\remains feasible\;
        \STATE Update the solution $\mathbf{x}_{k} \leftarrow \mathbf{x}_{k-1} + \alpha_k \Delta \mathbf{x}_k$\;
        \STATE Reduce the barrier parameter $\mu \leftarrow \tau \mu$\;
    \ENDWHILE
    \RETURN \!\!approximately optimal solution $\mathbf{x}_k$.
\end{algorithmic}
\end{algorithm}

The primary bottleneck of IPM when solving large-scale LP problems is the complexity of solving large linear systems of equations (the Newton's step in Line~\ref{l:newton}) at each iteration.
The fastest linear system solver runs in almost cubic time, significantly limiting the scalability of IPM beyond modest problem sizes.
Meanwhile, because of its use of the Newton's step, IPM can converge to optimality in almost a constant number of iterations in most practical settings, typically less than $5$-$20$ iterations regardless of problem size~\cite{ipm_25}.
This makes IPM a particularly good candidate to be mimicked and sped-up by ML---speeding-up the Newton's step will yield good asymptotic improvement, while the small number of iterations is ideal for algorithmic alignment, as we discuss in Sec.~\ref{sec:align}.%

\begin{figure}[!t]
\centering
{\includegraphics[width=0.42\textwidth]{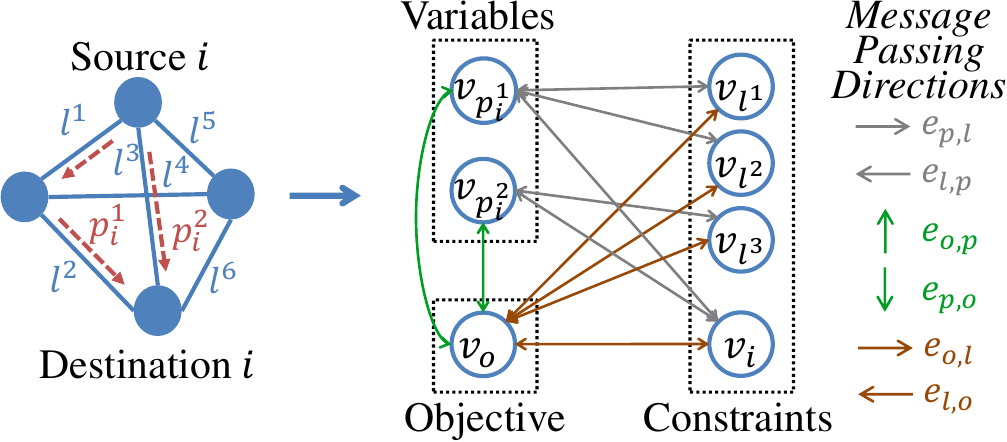}
}
\caption{Graph representation of TE instance with one SD pair and two paths. Each vertex denotes a variable, a constraint, or the objective. Directed edges denote their correlations in the LP formulation. GNN message passing will be carried out along the edges in six directions, as detailed in Sec.~\ref{sec:model_design}.
}
\label{fig:node_edge}
\end{figure}
\subsection{Representing TE as a Graph}
\vspace{-0.28em}

\noindent
To speed-up IPM, our first step is to properly represent the TE problem as a neural model that can be trained.
As we motivated in Sec.~\ref{sec:motivation}, a simple GNN defined over the network topology graph itself cannot effectively embed information such as decision variables (path traffic split ratios), constraints and the objective.
To this end, we develop a new GNN architecture, where the graph is not defined on the network but fully on the entities (variables, constraints and objective) and their correlations presented in the LP formulation in Program~\eqref{TE}.

Consider the GNN graph $G = (V, E)$ where $V$ is defined by four sets of vertices: set $V^{\sf p}$ denotes path vertices with $v_p \in V^{\sf p}$ for each $p \in P$; set $V^{\sf d}$ denotes SD pair vertices with $v_i \in V^{\sf d}$ for each SD pair $i \in D$; set $V^{\sf l}$ denotes link vertices with $v_l \in V^{\sf l}$ for each link $l$; and a single vertex $v_o$ denotes the network-wide objective function in Program~\eqref{TE}.
Each vertex in $V^{\sf p}$ represents a \emph{traffic splitting variable} $\mathcal{R}(p)$, each vertex in $V^{\sf d}$ represents a \emph{demand constraint}, and each vertex in $V^{\sf l}$ represents a \emph{capacity constraint}.

Edge set $E$ contains undirected edges that model correlations among path variables, link constraints, and the objective function.
The set $E^{\sf pd}$ contains edges $e_{p, i}$ that each models the correlation between path $p$ and SD pair $i$'s demands, for $p_i \in P_i$.
The set $E^{\sf pl}$ contains edges $e_{p, l}$ that each models the correlation between path $p$ and link $l$, if $l \in p$.
The set $E^{\sf po}$ contains edges $e_{p, o}$ that each models the correlation between path $p$ and objective $o$.
The set $E^{\sf do}$ contains edges $e_{i, o}$ that each models the correlation between SD pair $i$ and objective $o$.
The set $E^{\sf lo}$ contains edges $e_{l, o}$ that each models the correlation between link $l$ and objective $o$.
Intuitively, edges $E^{\sf pd}$ represent each path's traffic split ratio contribute to the demand of its SD pair, and edges $E^{\sf pl}$ represent how allocations on paths through the same link are bounded by the link capacity.
Edges $E^{\sf po}$ represent how each path's traffic split ratio contributes to maximizing the total throughput (the objective).
While edges $E^{\sf do}$ and $E^{\sf lo}$ may seem disconnected from Program~\eqref{TE}---there is no direct relationship between the constraints and the objective---they can in fact be interpreted as the implicit correlations between the \emph{dual variables} corresponding to the constraints, and the \emph{dual objective} which is equal to the primal objective by strong duality of LP.
Each edge is labeled with the corresponding coefficient in Program~\eqref{TE}.
For instance, edge $e_{p, i} \in E^{\sf pd}$ for path $p$ would have weight $1$ as the coefficient of variable $\mathcal{R}(p)$ in the demand constraint, and edge $e_{p, l} \in E^{\sf pl}$ for $p \in P_i$ would have weight $d_i$ as in the capacity constraint.

Fig.~\ref{fig:node_edge} shows the GNN graph representation for a network with four nodes and six links.
Given SD pair $i$, there are two possible paths $p_i^1$, $p_i^2$ (red dash lines), which are represented by the variable vertices: $v_{p_i^1}$ and $v_{p_i^2}$.
The capacity constraints for the three links $l^1$, $l^2$, $l^3$ are represented by constraint vertices $v_{l^1}$, $v_{l^2}$ and $v_{l^3}$.
$v_i$ represents the single demand constraint, and $v_o$ represents the objective.
%
%
The gray edges $e_{p, i}$ and $e_{p, l}$ models the relations between the variable vertices and the constraint vertices.
The green edges $e_{p, o}$ models the relations between the variable vertices and the objective vertex.
The brown edges $e_{i, o}$ and $e_{l, o}$  models the relations between the constraint vertices and the objective vertex.
Here, we represent this TE as a GNN.

Our approach differs from these works~\cite{teal, chen2022representing, ipmgnn, lee2020fast} in two ways. First, \cite{teal} uses GNN to learn the final solution pattern directly. It does not encode the LP structure of TE, so its generalization is limited. Second, \cite{chen2022representing, ipmgnn, lee2020fast} apply GNNs to a selected subset of LPs not including TE. 
Formulating and solving TE requires specialized domain knowledge, for instance, edge-flow versus path-flow formulation, network-agnostic representation of variables and constraints, and reaching network generalizability goals.
For \cite{chen2022representing}, we tested and observed weak feasibility and poor generalizability to large network sizes due to simple GNN architecture design.
We propose several key innovations, including path-flow formulation and neural representation of TE, dual-loop GNN architecture for generalizability, and TE-specific generalization-oriented training. We pretrain on small networks and evaluate on unseen instances. The model remains near-optimal on topologies up to $20\times$ larger (OGap $<\!4\%$).

\begin{figure}[t]
\centering
{\includegraphics[width=0.48\textwidth]{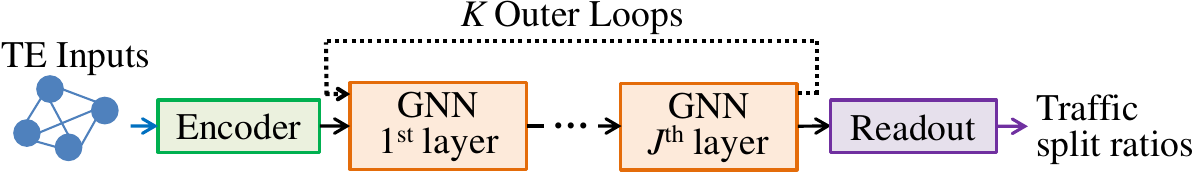}
}
\caption{Double-looped GNN architecture: $K$ outer loops matching (expected) number of IPM iterations, and $J$ inner layers.
}
\label{fig:ipmgnn}
\end{figure}
\subsection{Algorithmic Alignment and Strongly Supervised Training}
\label{sec:align}

\noindent
While existing work such as~\cite{teal, ipmgnn} utilized similar GNN structures as above, they trained their model to directly optimize for the objective function, regularized by constraint penalty.
This may however lead to difficulty in convergence, or converging to local optima that are far from global optimality.

Motivated by the algorithmic alignment in ML~\cite{chen2022representing, gnn_reason},
we want to better align the GNN model with the IPM algorithm to achieve strong performance and generalizability.
Since an IPM iteration basically takes a Newton's step from the previous iteration's solution, starting from an initial feasible solution, we want the GNN model to learn the same behavior.
To this end, we define a \textbf{double-looped} GNN architecture, which consists of $K$ outer GNN layers, and $J$ inner GNN layers in each outer layer, as in Fig.~\ref{fig:ipmgnn}.
Each outer layer corresponds to an IPM iteration, and seeks to mimic one Newton's update on the previous layer's solution (represented by vertex attributes after previous layer).
The number of outer layers, $K$, should correspond to the number of iterations that IPM takes to converge, which is usually a constant that can be easily estimated~\cite{ipmgnn}.

Because each outer layer corresponds to the same operation (Newton's step) that is applied on a different input (previous iteration's solution), the message passing and aggregation functions $g^k$ and $MSG^k$ in all outer GNN layers should share the same neural network parameters.
Furthermore, since the IPM generates a feasible solution in every iteration, each GNN outer layer's output can be supervised by a loss function defined between the GNN's output solution (by applying the readout function $r$) and the IPM's updated solution.
This \textbf{strong supervision} of each GNN layer's output can significantly improve the model's ability to learn not just the final optimal solutions in the training dataset, but also the \emph{trajectories of convergence to optimality} by IPM.
Such an ability is crucial for the neural model to learn the generalizability of the optimal algorithm---to converge to optimality given \emph{any} problem instance.

One concern in the above architecture is whether one (outer) GNN layer can well approximate the Newton's step, which is a fairly complex operation that warrants a nearly cubic complexity.
To this end, we further expand each outer layer to consist of $J$ inner GNN layers.
Intuitively, these layers are to learn the intermediate steps in solving a system of linear equations plus applying the system's solution to the variables' update.
Since these steps do not correspond to feasible solutions to the original problem, the inner GNN layers are unsupervised.
In practice, we find that $J=2$ inner layers are already sufficient for even the largest networks that we evaluate.

\subsection{Model Design and Training Procedures}
\label{sec:model_design}

\noindent
With the above discussions, we now describe the GNN model and training procedures for solving large-scale TE problems with generalizability as depicted in Fig.~\ref{fig:ipmgnn}.
%
%
The GNN model consists of $J$ sets of message passing and aggregation functions $\bm{MSG}_j$ and $\bm{g}_j$ corresponding to the $J$ inner layers, and one readout function $r$.
Each $\bm{MSG}_j$ contains six distinct functions: ${MSG}_j^{\sf p\text{-}dl}$, ${MSG}_j^{\sf dl\text{-}p}$, ${MSG}_j^{\sf p\text{-}o}$, ${MSG}_j^{\sf o\text{-}p}$, ${MSG}_j^{\sf dl\text{-}o}$, and ${MSG}_j^{\sf o\text{-}dl}$.
Each $\bm{g}_j$ contains three functions: ${g}_j^{\sf p}$, ${g}_j^{\sf dl}$, and ${g}_j^{\sf o}$.
For $\alpha, \beta \in \{ {\sf p, dl, o} \}$, the functions ${MSG}_j^{\sf \alpha\text{-}\beta}$ and ${g}_j^{\sf \beta}$ are used to update the vertex attributes of $V^{\beta}$ based on the attributes of $V^{\alpha}$, where ${\sf p}$ denotes path vertices, ${\sf dl}$ denotes demand and link vertices, and ${\sf o}$ denotes the objective vertex.
We cluster demand and link vertices together since all constraints can be normalized to have the same right-hand-side, \emph{i.e.}, let $\mathbf{b} = 1$ in the LP formulation.
All functions are shared across all outer layers, and the model can have a variable number of outer layers depending on the input instance.

The initial vertex attributes are denoted by $h_\text{p}^{0, 0}$, $h_\text{dl}^{0, 0}$ and $h_\text{o}^{0, 0}$ respectively.
These can be either fixed values, or problem-specific such as the mean and variance of each row/column.
Each GNN layer (outer $k$ inner $j$) then updates vertex attributes based on the update rules below:
\vspace{0.05in}
\begin{equation*}
\begin{aligned}
\label{inner}
    h_{\text{dl}}^{k, j} = & g_j^{\text{dl}} \Big( h_{\text{dl}}^{k, j-1}, MSG_j^{\text{p-dl}} (\{ h_{\text{p}}^{k, j-1} | e_{p, i} \in E^{\text{pd}}, e_{p, l} \in E^{\text{pl}} \}),  \\
    &  MSG_j^{\text{o-dl}} (\{h_{\text{o}}^{k, j-1} | e_{i, o} \in E^{\text{do}}, e_{l, o} \in E^{\text{lo}}\}) \Big);  \\
    h_\text{o}^{k, j} = & g_j^{\text{o}} \Big( h_{\text{o}}^{k, j-1}, MSG_j^{\text{dl-o}} (\{h_{\text{dl}}^{k, j} | e_{i, o} \in E^{\text{do}}, e_{l, o} \in E^{\text{lo}} \}),  \\
    &  MSG_j^{\text{p-o}} (\{h_\text{p}^{k, j-1} | e_{p, o} \in E^{\text{po}} \Big);  \\
    h_\text{p}^{k, j} = & g_j^{\text{p}} \Big( h_\text{p}^{k, j-1}, MSG_j^{\text{dl-p}} (\{h_{\text{dl}}^{k, j} | e_{p, i} \in E^{\text{pd}}, e_{p, l} \in E^{\text{pl}} \}),  \\
    &  MSG_j^{\text{o-p}} (\{h_\text{o}^{k, j} | e_{p, o} \in E^{\text{po}} \}) \}) \Big).  \\
\end{aligned}
\end{equation*}
Each outer iteration stacks $L$ graph-convolution layers aligned with IPM updates.
To stabilize deep propagation and mitigate over-smoothing, we adopt residual connections
\cite{li2020deepergcnneedtraindeeper,xu2018representationlearninggraphsjumping} in our model.

To train the GNN model, the training dataset (of size $M_{\sf tr}$) contains real or synthetic TE instances, each defined by a network topology and a set of SD pairs with demands.
An IPM algorithm is used to first solve each training instance to optimality, and the intermediate solutions are saved as part of the training dataset to supervise training.
Let $\mathcal{\widehat R}^{(\tau)}_k$ be the $k$-th IPM iteration's output solution when processing training instance $\tau$, and $\mathcal{R}^{(\tau)}_k$ be the solution output by applying the readout function $r$ to each variable vertex's attribute in $V^{\sf p}$.
The training loss function consists of three loss components:

(1) The \emph{variable loss} is defined as:
\begin{equation*}
\mathcal{L}_{\sf p} = \frac{1}{M_{\sf tr}} \sum_{\tau=1}^{M_{\sf tr}} \sum_{k=1}^{K_{\tau}} \xi^{K_{\tau}-k} \cdot \left| \mathcal{R}^{(\tau)}_k - \mathcal{\widehat R}^{(\tau)}_k \right|_2^2,
\end{equation*}
where $K_{\tau}$ is the number of IPM iterations (and hence number of outer loops) for instance $\tau$, and $\xi$ is a discount factor for each iteration to prioritize convergence in the later iterations.

(2) The \emph{constraint loss} is defined as:
\begin{equation*}
\begin{aligned}
\mathcal{L}_{\sf dl} 
&= \frac{1}{M_{\sf tr}} \sum_{\tau=1}^{M_{\sf tr}} \sum_{k=1}^{K_\tau} \xi^{K_\tau-k} \cdot \left(  \sum_{i \in D^{(\tau)}} \text{ReLU} \left(  f^D_i(\mathcal{R}_k^{(\tau)}) \right) \right. \\
&\quad 
+ \left.  \sum_{l \in L^{(\tau)}} \frac{\text{ReLU} \left(  f^L_l(\mathcal{R}_k^{(\tau)}) \right)}{C(l)}  \right),
\end{aligned}
\end{equation*}
where $\text{ReLU}$ is a rectified linear unit, and $D^{(\tau)}$ and $L^{(\tau)}$ are the corresponding demand set and link set in problem instance $\tau$.
While the IPM trajectory is feasible by construction, a learned GNN does not inherit these guarantees. Relying on the objective loss alone can produce infeasible solutions during intermediate iterations, which leads to either suboptimal convergence or infeasible final solutions. We therefore add an explicit feasibility signal at each outer loop.
The constraint loss is defined as the constraint violation of solution generated in every GNN outer loop, normalized by the right-hand-side constant of each constraint.
It penalizes the output solution of each outer loop to ensure feasibility of the solution update trajectory, in line with the original IPM.

(3) Finally, the \emph{objective loss} is defined as:
\begin{equation*}
\mathcal{L}_{\sf o} = \frac{1}{M_{\sf tr}} \sum_{\tau=1}^{M_{\sf tr}} \sum_{k=1}^{K_{\tau}} \xi^{K_{\tau}-k} \cdot \left| f_0(\mathcal{R}^{(\tau)}_k) - f_0(\mathcal{\widehat R}^{(\tau)}_{K_{\tau}}) \right|_2^2.
\end{equation*}
The objective loss penalizes when the GNN generates solutions that deviate too much from the IPM's updates in each iteration.

With the above, the final training loss is defined as 
\vspace{-0.05in}
\begin{equation}
    \label{eq:loss}
    \mathcal{L} = \rho_1 \mathcal{L}_{\sf p} + \rho_2 \mathcal{L}_{\sf dl} + \rho_3 \mathcal{L}_{\sf o}.
\vspace{0.05in}
\end{equation}

\vspace{-0.7em}
\section{Performance Evaluation}
\label{sec:eval}
\vspace{-0.5em}
\subsection{Datasets}
\noindent 
To thoroughly evaluate the performance and generalizability of our algorithm, we generated four datasets for training and testing: two with randomly generated topologies, and two with practical, real-world topologies.
Each dataset has a training set and a test set.
Their statistics are summarized in Table~\ref{tab:data_info}. 

\begin{table*}[htb]
\small
\caption{Statistics of network topology sizes in each training and test dataset}
\label{tab:data_info}
\centering
\setlength\tabcolsep{2pt}
\renewcommand{\arraystretch}{1.5}
\begin{tabular}{|c|c|c|c|c|}
\hline
\textbf{Dataset}     & \textbf{Erd\"os-R\'enyi Training}   & \textbf{Waxman Training}  & \textbf{B4 Training} & \textbf{ASN Training}                   \\ \hline 
\# of nodes & 
20, 30, \dots, 100                & 200, 300, \dots, 800              & 12          & 217, 237                \\ \hline
\# of edges & $O(10^2)$, \dots, $O(7.9\times10^3)$     & $O(6\times10^2)$, \dots, $O(4\times10^3)$    & 38          & $O(6\times10^2)$                 \\ \hline \hline
\textbf{Dataset}     & \textbf{Erd\"os-R\'enyi Test}   & \textbf{Waxman Test}  & \textbf{B4 Test} & \textbf{ASN Test}                   \\ \hline 
\# of nodes & 200, 500, 1000, 2000          & 2000, 3000, 4000, 5000          & 12          & 553, 1739               \\ \hline
\# of edges & $O(2.8\times10^4)$, \dots, $O(3.6\times10^6)$ & $O(1.1\times10^4)$, \dots, $O(2.1\times10^4)$ & 38          & $O(1.6\times10^3)$, $O(9\times10^3)$ \\ \hline
\end{tabular}
\vspace{0.07in}
\end{table*}

\textbf{Erd\"os-R\'enyi (ER) Dataset:}
We generated random topologies based on the Erd\"os-R\'enyi model~\cite{erdos}.
For \emph{training set}, we generated networks with $\{ 20, 30, \dots, 100 \}$ nodes, each with a connection probability of $\{ 0.3, 0.4, \dots, 0.8 \}$.
This resulted in $9 \times 6 = 54$ topology configurations, and we generated $200$ instances for each configuration.
For each instance, $10$ random SD pairs were selected.
The link capacities and SD pair demands were both randomly generated in $[1000, 5000]$; the same default range applies to all other datasets below.
For \emph{test set}, we generated significantly larger networks with $\{ 200, 500, 1000, 2000 \}$ nodes respectively, with connection probability in $\{ 0.1, 0.2, \cdots, 0.8, 0.9 \}$.
Each of the 36 test configuration had $300$ instances with the same number of SD pairs and range for demands and capacities; we denote these $36$ test sets as $\text{ER}^{\sf test}_{n, q}$, where $n$ is the node number, $q$ is the connection probability.
The largest topology in training set had almost $8\times 10^3$ links, and the largest topology in test set had $3.6 \times 10^6$ links, two to three orders of magnitude larger than training.

\textbf{Waxman (WA) Dataset:}
Random topologies were generated based on the Waxman model~\cite{waxman}.
For \emph{training set}, we generated networks with $\{ 200, 300, \dots, 800 \}$ nodes with the default connection parameters $\alpha = 0.1$ and $\beta = \{0.2, 0.18, \dots, 0.08\}$\footnotemark{}.
\footnotetext{Here the $\alpha, \beta$ are Waxman parameters~\cite{waxman}, not those used in Sec.~\ref{sec:model_design}.}%
Each configuration had $800$ instances, and each instance had $20$ random SD pairs.
For \emph{test set}, we generated large networks that had $\{ 2000, 3000, 4000, 5000 \}$ nodes, with the default connection parameters $\alpha = 0.1$ and $\beta = \{0.03, 0.02, 0.015, 0.01\}$ respectively, each with $500$ instances.
We set these parameters so that the generated networks have similar node/link ratios as the real-world one (ASN).
We denote the test sets as $\text{WA}^{\sf test}_{n}$ for node number $n$.
The largest topology in training set had $4\times 10^3$ links, and the largest topology in test set had $2 \times 10^4$ links, five times larger than training.

\textbf{B4 and ASN Datasets:}
Two real-world network topologies were used for evaluation which were also used in existing work~\cite{teal}: Google's cloud-WAN topology B4~\cite{b4}, and Microsoft's AS-level Internet topology denoted as ASN~\cite{asn}.
B4 is a small topology with $12$ nodes and $38$ links.
Thus we trained and tested on the same topology, with $200$ and $50$ training and test instances respectively.
ASN is a large topology with $1739$ nodes and $8558$ links.
For ASN, we chose two randomly sampled subgraphs as training set, with $217$ and $237$ nodes and each around $600$ links respectively; $2000$ instances were generated for each subgraph.
We tested on the largest connected component of ASN \emph{after pruning all nodes in the training set}, with $553$ nodes denoted as $\text{ASN}^{\sf test}_{553}$ and around $1600$ links, as well as the full ASN topology denoted as $\text{ASN}^{\sf test}_{1739}$; each had $500$ test instances.
%
%
The largest test instance in ASN had more than $10\times$ links than the largest training instance. 
\noindent 
\subsection{Implementation and Evaluation Setup}
\noindent 
\textbf{Comparison Algorithms:}
We compared TELGEN to two state-of-the-art learning-based TE solutions, TEAL~\cite{teal} and HARP~\cite{trans_wan}, and also the optimal SciPy IPM solver as the baseline.\footnote{In addition to these, we have also tested Google's OR-Tools~\cite{cpsatlp} open-source LP solver. Unfortunately, OR-Tools cannot generate feasible solutions consistently as the other solvers, and fails to solve many instances when network size becomes large (\emph{e.g.} $n \ge 1000$). Hence we omit the comparison results with OR-Tools below.}
Below, we describe their implementation details.
\subsubsection{TELGEN}
We extended the GNN codebase in~\cite{ipmmpnn_git} to implement our model.
The GNN model was composed of MLP encoders each to encode the initial attributes of one set of vertices in $\{ V^{\sf p}, V^{\sf d} \cup V^{\sf l}, \{ v_o \} \}$, a series of graph convolutional (GCN) layers to perform messaging among vertices, and an MLP readout function to generate the TE solution from each path vertex's attribute in $V^{\sf p}$; a ReLU layer was applied after the readout function to ensure non-negativity of the TE solution.
%
%

All encoders were all two-layer MLPs with hidden dimensions 180 and 360.
We used $J=2$ GCN layers for the inner loop. 
For ER and B4 datasets, we used $8$ outer layers; for Waxman and ASN, we used $16$ outer layers.
The hidden dimensions of the readout two-layer MLP were 360 and 720. 
The model was trained independently for $150$ epochs on ER and WA datasets, and $50$ epochs on the B4 and ASN datasets.
\subsubsection{TEAL~\cite{teal}}
We used the open-source implementation of TEAL with the original model architecture, hyperparameters and same training epochs as TELGEN to make fair comparison.
As a remark, TEAL is designed in a way such that testing must be on the same topology and SD pairs as training.

{%

\subsubsection{HARP~\cite{trans_wan}}
We used the open-source implementation of HARP with the original
model architecture, 
hyperparameters and same training epochs as TELGEN to make fair comparison.
One issue of HARP is that it has a large RAM footprint on large topologies, and even our latest hardware platform (as below) could not execute HARP on the largest topology---the full ASN.
To make a fair comparison between TELGEN and the two baselines, we trained and tested all three algorithms above on B4 and a randomly sampled subnetwork of ASN in lieu of the full ASN topology.
}

\noindent 
\subsubsection{SciPy~\cite{SciPy}}
The {\small\sf scipy.optimize.linprog} routine was used to optimally solve the LP with IPM, and the per-iteration solution was recorded to supervise training of TELGEN.

\subsubsection{Gurobi~\cite{gurobi}}
We used the Gurobi Optimizer to obtain ground-truth optimal objectives for gap computation and also recorded its wall-clock solution time under default settings (LP barrier/simplex as chosen by the solver), without warm-starts.%
\noindent
\textbf{Hardware platform:}
The algorithms were implemented and evaluated on a desktop server with {Intel Xeon Gold 5317 CPU (3.0GHz, 48 cores), 256GB RAM and Nvidia A100 GPU (40GB).}
All experiments were run with $3$ random seeds and averages are shown below.
\paragraph{Runtime accounting}
We report training time and per-instance inference latency instead of memory.
TELGEN incurs a one-time training cost (about 4h), after which inference is millisecond-level.
In contrast, classical solvers re-solve every instance from scratch.
For a batch of $N$ instances, the amortized time is
$T_{\text{TELGEN}}=T_{\text{train}}+N\cdot t_{\text{inf}}$ versus $T_{\text{solver}}=N\cdot t_{\text{solve}}$,
favoring TELGEN for moderate $N$.

\noindent 
\subsection{Evaluation Metrics}

\noindent
We consider five metrics for comparison among algorithms.
Consider a test set with $M_{\sf ts}$ instances.
The \emph{objective gap (\textbf{OGap})} is defined as the mean absolute difference between the objective value of the final algorithm output solution $\mathcal{\widehat R}^{(\tau)}$ and the optimal solution $\mathcal{\widehat R}^{(\tau)}$ (output by the IPM solver) for test instant $\tau$, normalized by the optimal value: 
\begin{equation}
    \gamma_{\sf obj} \triangleq \frac{1}{M_{\sf ts}} \sum_{\tau=1}^{M_{\sf ts}} \left|\frac{ f_0(\mathcal{R}^{(\tau)}) - f_0(\mathcal{\widehat R}^{(\tau)} )}{f_0(\mathcal{R}^{(\tau)})}\right|. 
\end{equation}
The \emph{constraint gap (\textbf{CGap})} is the mean absolute constraint violation of the final algorithm output solution: 
\vspace{-0.5em}
\begin{equation}
\begin{aligned}
\gamma_{\sf con} 
&= \frac{1}{M_{\sf ts}} \sum_{\tau=1}^{M_{\sf ts}} \left(  \sum_{i \in D^{(\tau)}} \text{ReLU} \left(  f^D_i(\mathcal{R}^{(\tau)}) \right) \right. \\
&\quad 
+ \left.  \sum_{l \in L^{(\tau)}} \frac{\text{ReLU} \left(  f^L_l(\mathcal{R}^{(\tau)}) \right)}{C(l)}  \right).
\end{aligned}
\vspace{-0.5em}
\end{equation}
It is similar to the constraint loss but only considers final output of the algorithm.
The \emph{objective gap without constraint violation (\textbf{OnoCGap})} is the objective gap when the solution is scaled by the maximum ratio of any constraint violation, thereby ensuring the final solution is feasible.
Finally, the \emph{\textbf{training time}} per epoch and the \emph{\textbf{prediction time}} measure the time spent on training for each epoch on average, and the time for making each prediction on average during testing (or running IPM on one instance till optimality for SciPy), respectively.

\noindent 
\subsection{Evaluation Results}
\begin{table}[t]
\small
\caption{\textbf{OnoCGap} (\%) for TEAL and TELGEN}
\centering
\setlength\tabcolsep{1pt}
\renewcommand{\arraystretch}{1.5}
\label{onocgap_teal_ipmgnn}
\begin{tabular}{|P{2.2cm}|P{1.8cm}|P{1.8cm}|P{1.8cm}|}
\hline
\textbf{OnoCGap (\%)} & \textbf{HARP} & \textbf{TEAL}    & \textbf{TELGEN} \\ \hline
B4  & 26.99 & 26.21 & 1.75 \\ \hline
$\text{ASN}_{98}$ & 73.04 & 98.07 & 1.46 \\ \hline
\end{tabular}
\end{table}
\begin{table}[t]
\small
\caption{Model training time per epoch and prediction time per test instance}
\centering
\setlength\tabcolsep{1pt}
\renewcommand{\arraystretch}{1.5}
\label{prac_train_test_time}
\begin{tabular}{|P{3.5cm}|P{1.5cm}|P{1.5cm}|P{1.5cm}|}
\hline
\textbf{Training Time (second)} & \textbf{HARP}
&\textbf{TEAL}    & \textbf{TELGEN}   \\ \hline 
B4   &  0.86  & 2.00    & 1.63    \\ \hline
$\text{ASN}_{98}$  &  12.59  & 7.80 & 2.56    \\ \hline \hline
\textbf{Prediction Time (ms)} & \textbf{HARP}  &\textbf{ TEAL}  & \textbf{TELGEN} \\ \hline 
B4   &  29.28  & 5.04 & 1.55    \\ \hline
$\text{ASN}_{98}$  &  814.27
  &  4.5 & 1.52    \\ \hline
\end{tabular}
\end{table}
\noindent 
\subsubsection{Comparing TELGEN,  TEAL and HARP}
{%
We first compared the performance of TELGEN to TEAL and HARP, the state-of-the-art learning-based TE solutions.
Since TEAL needs to be trained and tested on the same network topology while HARP cannot process topology as large as the whole ASN graph, 
we evaluated all three algorithms by training and testing them on datasets generated on B4 and a connected sub-network of ASN\footnotemark{} with $98$ nodes and $202$ egdes.
\footnotetext{These are the smallest and largest topologies evaluated in~\cite{teal} respectively, though we train and test on different SD pairs and demands than original paper.}
Table~\ref{onocgap_teal_ipmgnn} shows \textbf{OnoCGap}---the objective gap when all solutions are rounded to be feasible---of TELGEN, TEAL and HARP on the two datasets.
To our surprise, TEAL performed poorly even on the small B4 topology, achieving as high as $26\%$ objective gap; on the sub-ASN topology, it had a $98\%$ objective gap.
HARP performed almost as badly as TEAL as shown: it has a $27\%$ objective gap on B4 and $73\%$ on sub-ASN.
As we investigate, the reason for TEAL's inferior performance was due to lack of SD pairs for building the traffic matrix in the training dataset.
The reasons for HARP's performance were
1) the different objective of HARP leading to a different model architecture: HARP's model architecture is designed to meet a specific goal which is maximum link utilization (MLU) in the paper while our TELGEN focuses on generalization across diverse topologies and network conditions, thus suits for different objectives;
2) In order to optimize for the MLU, HARP tended to allocate data flow through the few same nodes which caused bottleneck links and lead to the poor performance;
3) lack of SD pairs for building the traffic matrix in the training dataset which is similar as TEAL.
Compared to TEAL and HARP, TELGEN achieved negligible ($<3\%$) objective gap, and even performed better on the  larger sub-ASN topology with $1.7\%$ gap after rounding.

Though HARP showed in~\cite{trans_wan} that it could handle topology variations, it only allowed $20\%$ of links capacity variation with ratio of less than 0.8, with about $5\%$ of the links with zero capacity (considering as inactive links) while TELGEN allows completely different topologies for training and testing.
When considering topology changes for training and testing dataset, HARP only showed the effectiveness using AnonNet dataset with different sets of active nodes as clusters containing dozens of nodes but no larger graphs containing thousands or even more nodes like TELGEN.
HARP also show the effectiveness in perturbations on a large topology, KDL with a 754 nodes, while TELGEN showed the generalization across different and much larger topology with the number of nodes up to $10^6$ as shown in the next section.
This further showed the superior of our algorithm TELGEN.
However, we are also aware that HARP is specially designed for an objective and the data we used might be not the best fit for HARP algorithm.
We shall implement and adjust the HARP dataset so as to fit our TELGEN as a future work.

Both TEAL and HARP had comparable or faster \textbf{training time} per epoch as TELGEN on B4, but both became at least $3\times$ slower than TELGEN when facing the larger sub-ASN topology as shown in Table~\ref{prac_train_test_time}.
This was due to TEAL's fairly complex model architecture which combines GNN with MARL, and even an expensive classical ADMM algorithm that must be run on CPU during end-to-end training.
HARP's model architecture could be even more complex that it starts with a GNN layer, after which comes a transformer as well as an MLP, then ended with recurrent adjustment unit (RAU) consisting layers of MLPs.
TELGEN was faster because of its simple GNN architecture, and its entire training can be conducted on GPU.
The reason that TEAL and HARP were slower than TELGEN is also because they both consider demands of all possible SD pairs among all nodes in the network, which means they take much larger size of inputs than TELGEN.
On the contrary, TELGEN can be applied to an arbitrary number of SD pairs in the network which shows the flexibility of our method.
For \textbf{prediction time} when making a prediction, TELGEN again significantly outperformed TEAL by over $3\times$ for B4 and sub-ASN, and HARP by over $18\times$ for B4 and $500\times$ for sub-ASN due to its simple, GPU-compatible architecture.
The significant superior performance of TELGEN for \textbf{training time} and \textbf{prediction time} show that our model can be generalized to different SD pairs as well as topology sizes and be used with flexibility. 

While the above results show TELGEN's superior performance and efficiency compared to state-of-the-art algorithm when training and testing on the same topology, the next sets of experiments fully unveil TELGEN's generalization power, for which no existing learning-based TE solution can compare.

\noindent 
\subsubsection{Generalization on Real-world Networks}
We next show TELGEN's generalization power on real-world networks, and compare its performance to the other generalizable baseline---the SciPy IPM solver itself.
The results are shown in Table~\ref{tab:b4_asn_gap}.
\begin{table}[t]
\small
\caption{B4 and ASN: gaps (\%) and prediction time (ms)}
\label{tab:b4_asn_gap}
\centering
\setlength\tabcolsep{1pt}
\renewcommand{\arraystretch}{1.5}
\begin{tabular}{|P{1.2cm}|P{1.4cm}|P{1.4cm}|P{1.4cm}|P{1.4cm}|P{1.4cm}|}
\hline
\multirow{2}{*}{\textbf{Dataset}} & \multicolumn{3}{c|}{\textbf{Gaps (\%)}} & \multicolumn{2}{c|}{\textbf{Time (ms)}} \\ 
\cline{2-6}
& \textbf{OGap} & \textbf{CGap} & \textbf{OnoCGap}  & \textbf{TELGEN} & \textbf{SciPy}\\ \hline
$\text{B4}$ & 5.94 & 1.62  & 1.75  & 1.55 & 9.41\\ \hline
$\text{ASN}^{\sf test}_{553}$ & 7.53 & 3.57  & 1.67  & 3.70 & 12.84 \\ \hline
$\text{ASN}^{\sf test}_{1739}$ & 12.25 & 5.68 & 2.00 & 3.73 & 11.43 \\ \hline
\end{tabular}
\end{table}
The B4 results were the same as in Table~\ref{onocgap_teal_ipmgnn}.
For the two ASN test sets, note that the training set contained only two \textbf{much smaller subnetworks} of the entire network, with around $40\%$ of the links as in $\text{ASN}^{\sf test}_{553}$ and $7\%$ of the links as in $\text{ASN}^{\sf test}_{1739}$.
Also note that the training set had \textbf{no overlap} in nodes or links with the $\text{ASN}^{\sf test}_{553}$ due to how we constructed the training and test sets.
Nevertheless, we observe that TELGEN achieved almost negligible OGap ($<\!\!2.5\%$) and CGap ($<\!\!0.65\%$) in both test sets compared to the optimal solution output by IPM.
After scaling, the \textbf{OnoCGap} was $<\!\!2.1\%$.
The OnoCGap was \emph{reduced} compared to OGap since TELGEN tended to ``over-provision''---allocating more than link capacities which led to (infeasibly) higher objective than the optimal solution---which was corrected after the solution was scaled down by capacity violation.
This was exactly the behavior that we wanted TELGEN to learn: to maximize the objective as the IPM algorithm did even when the ``maximize'' symbol itself was not encoded in the GNN inputs at all.

For \textbf{prediction time}, TELGEN achieved $16.5\%$-$71.2\%$ reduction over SciPy.
The advantage is expected to become larger with larger TE instances (e.g.\ larger networks and more SD pairs).
In principle, TELGEN involves linear complexity in the number of variables while the fastest LP solver involves almost cubic complexity.
Moreover, TELGEN can utilize the massive parallelism enabled by GPU, whereas traditional solvers can at best perform CPU parallelism, further limiting their scalability.

\begin{table*}[htb]
\small
\caption{Gaps (\%) for ER test datasets across varying densities}
\label{tab:erdos_renyi_gap}
\centering
\setlength\tabcolsep{4pt} 
\renewcommand{\arraystretch}{1.2}
\begin{tabular}{|l|c|c|c|c|}
\hline
\multicolumn{1}{|c|}{\textbf{Dataset}} & \textbf{OGap (\%)} & \textbf{CGap (\%)} & \textbf{OnoCGap (\%)}  & \textbf{Avg\_inference\_time (ms)}\\ \hline
$\text{ER}^{\sf test}_{200, 0.1}$ & 0.88 & 6.14 & 2.30  & $2.43 \pm 0.87$ \\ \hline
$\text{ER}^{\sf test}_{200, 0.5}$ & 0.05 & 4.43 & 0.25  & $1.98 \pm 1.26$\\ \hline
$\text{ER}^{\sf test}_{200, 0.9}$ & 0.08 & 6.76 & 0.32  & $1.95 \pm 1.27$\\ \hline
$\text{ER}^{\sf test}_{500, 0.1}$ & 0.13 & 5.52 & 0.69  & $2.00 \pm 1.26$\\ \hline
$\text{ER}^{\sf test}_{500, 0.5}$ & 0.07 & 3.44 & 0.19  & $1.92 \pm 1.29$\\ \hline
$\text{ER}^{\sf test}_{500, 0.9}$ & 0.08 & 5.11 & 0.25  & $1.97 \pm 1.28$\\ \hline
$\text{ER}^{\sf test}_{1000, 0.1}$ & 0.07 & 3.68 & 0.22  & $1.90 \pm 1.24$\\ \hline
$\text{ER}^{\sf test}_{1000, 0.5}$ & 0.06 & 3.04 & 0.18  & $1.88 \pm 1.88$\\ \hline
$\text{ER}^{\sf test}_{1000, 0.9}$ & 0.07 & 3.16 & 0.18  & $1.92 \pm 1.31$\\ \hline
$\text{ER}^{\sf test}_{2000, 0.1}$ & 0.08 & 3.68 & 0.21  & $1.87 \pm 1.22$\\ \hline
$\text{ER}^{\sf test}_{2000, 0.5}$ & 0.09 & 3.83 & 0.22  & $1.97 \pm 1.28$\\ \hline
$\text{ER}^{\sf test}_{2000, 0.9}$ & 0.05 & 1.49 & 0.10  & $1.91 \pm 1.22$\\ \hline
\end{tabular}
\end{table*}

\begin{table}[htb]
\small
\caption{Gaps (\%) for all WA test datasets}
\label{tab:wa_gap}
\centering
\setlength\tabcolsep{1pt}
\renewcommand{\arraystretch}{1.5}
\begin{tabular}{|>{\raggedright}P{1.4cm}|P{1.8cm}|P{1.8cm}|P{1.8cm}|}
\hline
\multicolumn{1}{|c|}{\textbf{Dataset}} & \textbf{OGap} & \textbf{CGap} & \textbf{OnoCGap}  \\ \hline
$\text{WA}^{\sf test}_{2000}$ & 2.08 & 0.46   & 1.10  \\ \hline
$\text{WA}^{\sf test}_{3000}$ & 2.16 & 0.47 & 1.16  \\ \hline
$\text{WA}^{\sf test}_{4000}$ & 2.21 & 0.46 & 1.20 \\ \hline
$\text{WA}^{\sf test}_{5000}$ & 3.70   & 0.62 & 1.89  \\ \hline
\end{tabular}
\end{table}
\subsubsection{Generalization on Random Topologies}
 In random dataset experiments, we trained our TELGEN model on the 54 topology configurations for ER dataset and 7 topology configurations for WA dataset.
The trained models had been tested on 12 test graphs with different number of nodes and probability for ER graphs while 4 test graphs with different number of nodes and $\beta$ for WA graphs.
Table~\ref{tab:erdos_renyi_gap} and~\ref{tab:wa_gap} show the test dataset detailed node number and edge information as well as three kinds of gaps: \textbf{OGap}, \textbf{CGap}, \textbf{OnoCGap} for each test dataset.
For both ER and WA datasets, TELGEN achieved negligible objective and constraint gaps, and the OnocGap was below $<\!\!2\%$ for all test sets.
The increase in network sizes in the test sets led to negligible impact on the gaps as well, demonstrating strong generalizability of TELGEN even when the test networks had $2$-$20\times$ more nodes than the largest training networks.
The WA test set witnessed slightly larger gaps when the network became very larger with $6\times$ more nodes than training, but the result was still superior.

Fig.~\ref{inference_errorbars_boxplot} reports the distribution of inference time across edge probability $p \in [0.1, 0.9]$ and graph sizes $n \in \{ 200,500,1000,2000 \}$.
Among the remaining solvers, TELGEN (teal) consistently achieves a lower mean latency than both \textsc{Scipy} and \textsc{Gurobi}, while maintaining tight variability across all $p$ and $n$.
Its runtime grows smoothly with $n$ and shows weak sensitivity to $p$, reflecting TELGEN’s fixed-depth, structure-aware forward pass; in contrast, classical LP/MIP pipelines require per-instance linear solves or search, resulting in higher averages runtime.
Fig.~\ref{fig:wa_time} shows that TELGEN again outperforms traditional solvers, and the advantage enlarges as the problem size (the number of SD pairs) increases.
When the SD pair count reaches $500$, TELGEN is over $200\times$ faster than the traditional \textsc{Scipy} solver, validating the asymptotic advantage (linear vs.\ cubic) and demonstrating significant efficiency for large-scale problems.

The advantage of TELGEN enlarged when the problem size---the number of SD pairs---increased as in Fig.~\ref{fig:wa_time}(b).
When the SD pair number went up to $500$, TELGEN sped up the traditional SciPy solver by over $200\times$, validating the asymptotic advantage (linear vs.\ cubic) and demonstrating significant efficiency improvement for large-scale problems. 

\begin{figure}
    \centering
    \includegraphics[width=1.1\linewidth]{img/inference_boxplot_30_24_new.jpg}
    \caption{Inference time (ms) across edge probability $p$ for ER graphs, faceted by size $n \in \{200,500,1000,2000\}$.}
    \label{inference_errorbars_boxplot}
\end{figure}

TELGEN completes a one-time training phase (around four hours) and subsequently performs inference within milliseconds.
By contrast, \textsc{Scipy} re-solves each instance without reuse of prior computations, resulting in significantly higher average computation delay and frequent outliers across edge probabilities and graph scales, as illustrated in Fig.~\ref{inference_errorbars_boxplot}.
\begin{figure}[htb]
\centering
\subfloat[20 SD pairs]
{\includegraphics[width=0.236\textwidth, trim=0 0 0 0, clip]{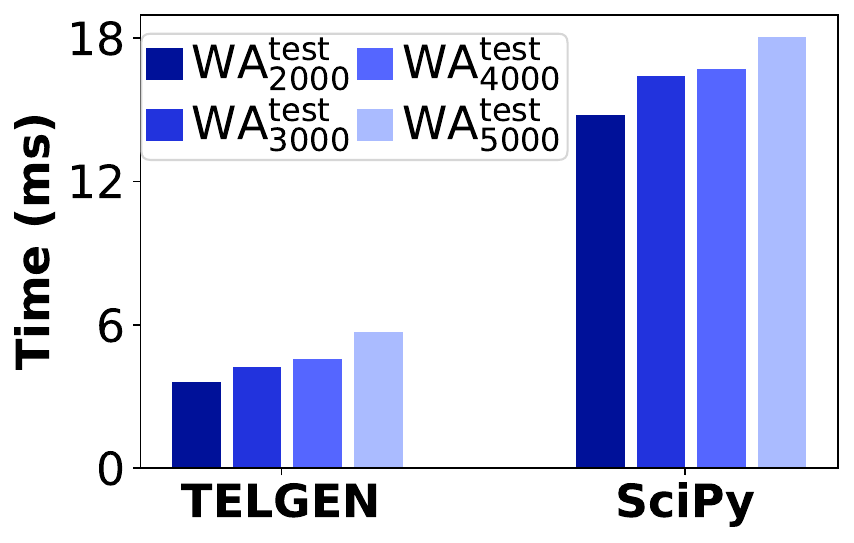}
}
\hfil
\subfloat[500 SD pairs]
{\includegraphics[width=0.236\textwidth, trim=0 0 0 0, clip]{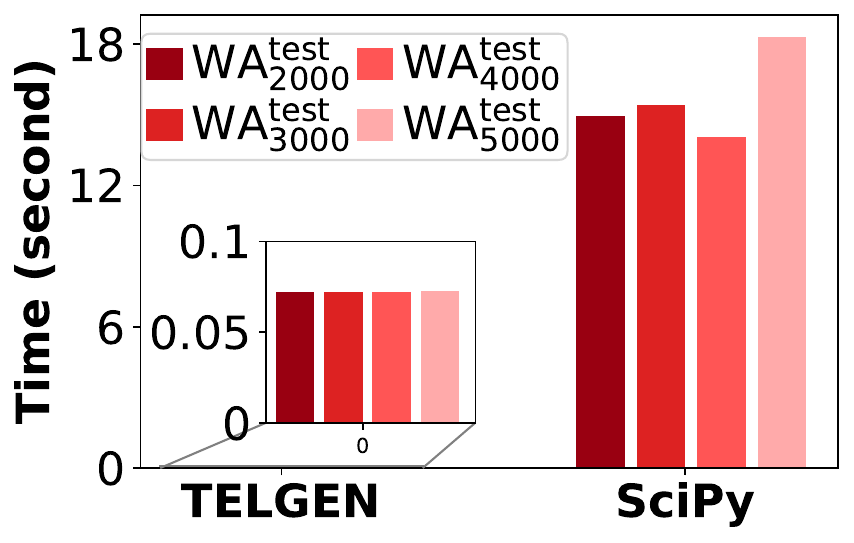}
}
\hfil
\caption{Prediction time for different WA test datasets based on number of nodes and different number of SD pairs.}
\label{fig:wa_time}
\end{figure}

\begin{table}[htb]
\small
\caption{ER Gaps (\%) with unseen demand distributions}
\label{tab:erdos_dmd_gap}
\centering
\setlength\tabcolsep{1pt}
\renewcommand{\arraystretch}{1.5}
\begin{tabular}{|>{\raggedright}P{1.4cm}|P{2.2cm}|P{1.4cm}|P{1.4cm}|P{1.4cm}|}
\hline
\multicolumn{1}{|c|}{\textbf{Dataset}} & \textbf{Demand range} & \textbf{OGap} & \textbf{CGap} & \textbf{OnoCGap} \\ \hline
$\text{ER}^{\sf test}_{200, 0.9}$ & [1200, 6000]&  0.88 & 6.14 & 2.30 \\ \hline
$\text{ER}^{\sf test}_{500, 0.9}$ & [1200, 6000]&  0.08 & 5.11 & 0.25 \\ \hline
$\text{ER}^{\sf test}_{1000, 0.9}$ & [1200, 6000]&  0.07 & 3.16 & 0.18 \\ \hline
$\text{ER}^{\sf test}_{2000, 0.9}$ & [1200, 6000]&  0.05 & 1.49 & 0.10 \\ \hline
$\text{ER}^{\sf test}_{200, 0.9}$ & [800, 4000]&  3.02 & 0.15 & 2.86 \\ \hline
$\text{ER}^{\sf test}_{500, 0.9}$ & [800, 4000]&  2.99 & 0.15 & 2.86 \\ \hline
$\text{ER}^{\sf test}_{1000, 0.9}$ & [800, 4000]&  3.01 & 0.15 & 2.85 \\ \hline
$\text{ER}^{\sf test}_{2000, 0.9}$ & [800, 4000]&  3.00 & 0.15 & 2.84 \\ \hline
\end{tabular}
\end{table}

\noindent 
\subsubsection{Generalization to Unseen Demand Distributions}
We evaluate TELGEN on Erd\H{o}s--R\'enyi (ER) test graphs of different scales and sparsities, using node counts $n\!\in\!\{200,500,1000,2000\}$ and connection probabilities $p$ from $0.1$ to $0.9$.
As reported in Table \ref{tab:erdos_renyi_gap}, TELGEN maintains consistently small errors across all settings: OGap ranges 0.05–0.88\%, CGap ranges 1.49–6.14\%, and the stricter OnoCGap remains within 0.10–2.30\%. 
The per-instance prediction time is low and stable, with means in 1.87–2.43\,ms and standard deviations 0.87–1.88\,ms, showing weak dependence on $n$ or $p$. 
As shown in Table \ref{tab:erdos_dmd_gap}, the gaps remain nearly identical across all unseen demand distributions, demonstrating TELGEN’s strong robustness to varying demand ranges. The OGap and OnoCGap values fluctuate within a narrow margin of ±0.02–0.03 \%, while CGap stays constant at 0.15 \%. Notably, the performance is invariant to both network size (200–2000 nodes) and demand scale ([800, 4000] vs. [1200, 6000]), confirming that TELGEN generalizes well to unseen traffic intensities without degradation in optimality.
These results indicate that TELGEN generalizes well across graph scale and density while preserving near-optimality and fast inference.

Summarizing the above results, we believe TELGEN can be particularly beneficial in two scenarios:
(1) A large-scale network operator can train the model on smaller subnetworks, such as within a region, while generalizing the trained model to the entire WAN without expensive re-training.
(2) A network software provider may train general TE models using small known or random topologies, and ship these models as part of their software to operators who only need slight fine-tuning to generalize these models onto their own, large-scale networks.

\section{Discussion and limitations}
\label{sec:DISCUSSION}

\noindent
In this work, we proposed TELGEN, a novel GNN-based framework for Traffic Engineering that achieves unprecedented generalizability and real-time performance. 
By learning the algorithmic process of IPM, TELGEN can be trained on small networks and deployed on topologies up to $20\times$ larger with near-optimal results. 
This addresses the critical scalability and deployment challenges faced by existing learning-based TE methods.
While our results are promising, it is important to discuss the limitations of our current work and the future research directions they inspire.

\textbf{Network Model Applicability:} Our work focuses on large-scale wired networks, such as data centers and cloud WANs, and on wireless backbones with point-to-point links, like those in LEO satellite constellations. However, it is not directly applicable to general wireless networks, like cellular or Wi-Fi, where link capacity is dynamic and affected by interference. To address the limitations in network scope, future work could explore new GNN architectures that model the dynamics of general wireless networks. 

\textbf{Dynamic Networks and Online Adaptation:} Extending TELGEN to explicitly handle dynamic networks is crucial for practical deployments. Our current model already provides a strong foundation: its generalizability across diverse topologies lets it adapt to discrete topological changes (e.g., link failures) \emph{without retraining}, and its millisecond-level prediction speed enables real-time reactions to rapidly changing traffic. At present we treat each time snapshot independently. Moving forward, we will incorporate temporal information and online adaptation mechanisms to leverage the continuous nature of network dynamics. These additions aim at a fully adaptive TE solution that proactively captures dynamic network conditions while preserving fast, stable inference.

\textbf{Distributed and Dynamic Design:} TELGEN currently operates in a centralized setting, assuming a global view of the network. It also treats each TE problem as an independent, static snapshot in time. This design does not explicitly model the continuous evolution of traffic demands and may face scalability or single-point-of-failure challenges in massive, decentralized environments. A key future direction is to develop a fully dynamic and distributed TELGEN. This enhanced model would use continuous temporal changes as features to make proactive decisions and employ decentralized agents that coordinate via other novel GNN architectures \cite{11085935,9414098,10.1145/3565473.3569185}, removing the need for a central controller.

\section{Conclusions}
\label{sec:conclusions}
\noindent
This paper proposed a method of utilizing machine learning to tackle the challenges of large-scale network TE problems and generalization to diverse network configurations.
TELGEN, a learning-based TE framework is designed to efficiently solve large-scale TE problems with generalizability.
Our framework employs modeling TE problems with general graphs, designs a novel double-looped GNN architecture to enable algorithmic alignment, and learns to generalize to unseen network scenarios by mimicking the step-by-step solving process of an optimal TE algorithm.
By capturing the relationships among problem variables, objectives, and constraints, TELGEN builds a structured graph. 
The framework is flexible, allowing for different configurations between training and testing networks by learning the intermediate outcomes of the interior-point method and maintaining input size and topology independence.
To evaluate its performance, we trained and tested TELGEN on large-scale randomly generated and real-world network topologies.
We demonstrated that TELGEN is suitable and generalizable for large-scale random and practical datasets by using different training and testing dataset configurations.
Extensive experiments have demonstrated that TELGEN can solve large-scale TE problems close to optimality with multi-fold improvement in prediction time compared to the optimal TE solver, reduce training time by one orders of magnitude over state-of-the-art learning-based TE solution, and generalize to unseen demand distributions and networks that are much larger than the training networks.
Our method outperformed the state-of-the-art learning-based method by prediction gap as well as efficiency.
It also outperformed traditional optimization method by efficiency in both small-scale and large-scale datasets.
Our proposed TELGEN resolved the scalability issue as well as the generalization challenge for wide-area network traffic engineering.

\bibliographystyle{myIEEEtranS}
\bibliography{IEEEabrv,ref}

\begin{IEEEbiography}[{\includegraphics[width=1in,height=1.25in,clip,keepaspectratio]{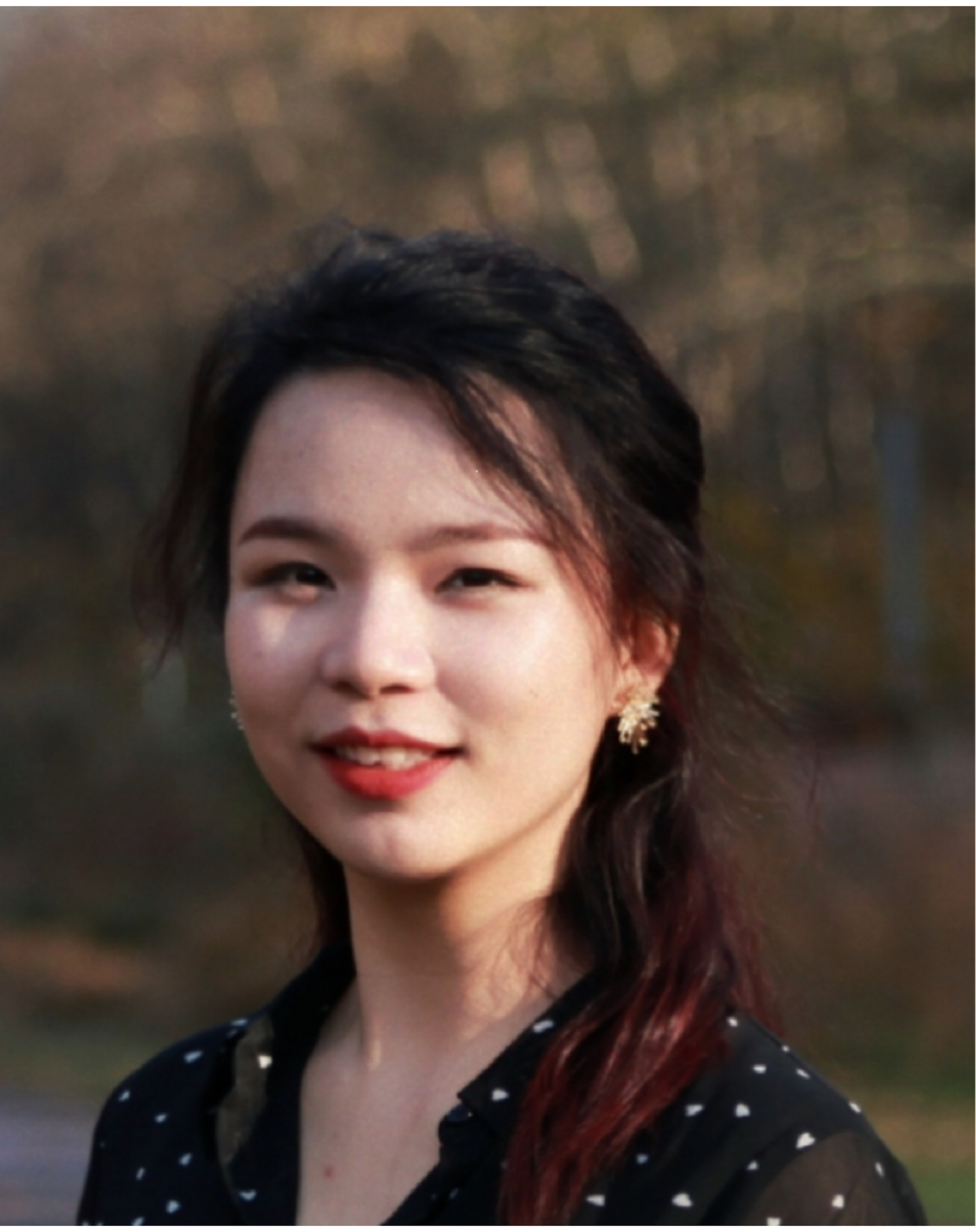}}]{Fangtong Zhou}(Student Member 2021) received her B.E. degree (2018) in Electrical Engineering and Automation from Harbin Institute of Technology, Harbin, China and M.S. degree (2020) in Electrical Engineering from Texas A\&M University, College Station, Texas, USA. She has been a Ph.D candidate in the Department of Computer Science at North Carolina State University while working on this paper. Her research interests include machine learning in computer networking, such as federated learning, reinforcement learning for resource provisioning. 
\end{IEEEbiography}
\begin{IEEEbiography}
[{\includegraphics[width=1in,height=1.25in,clip,keepaspectratio]{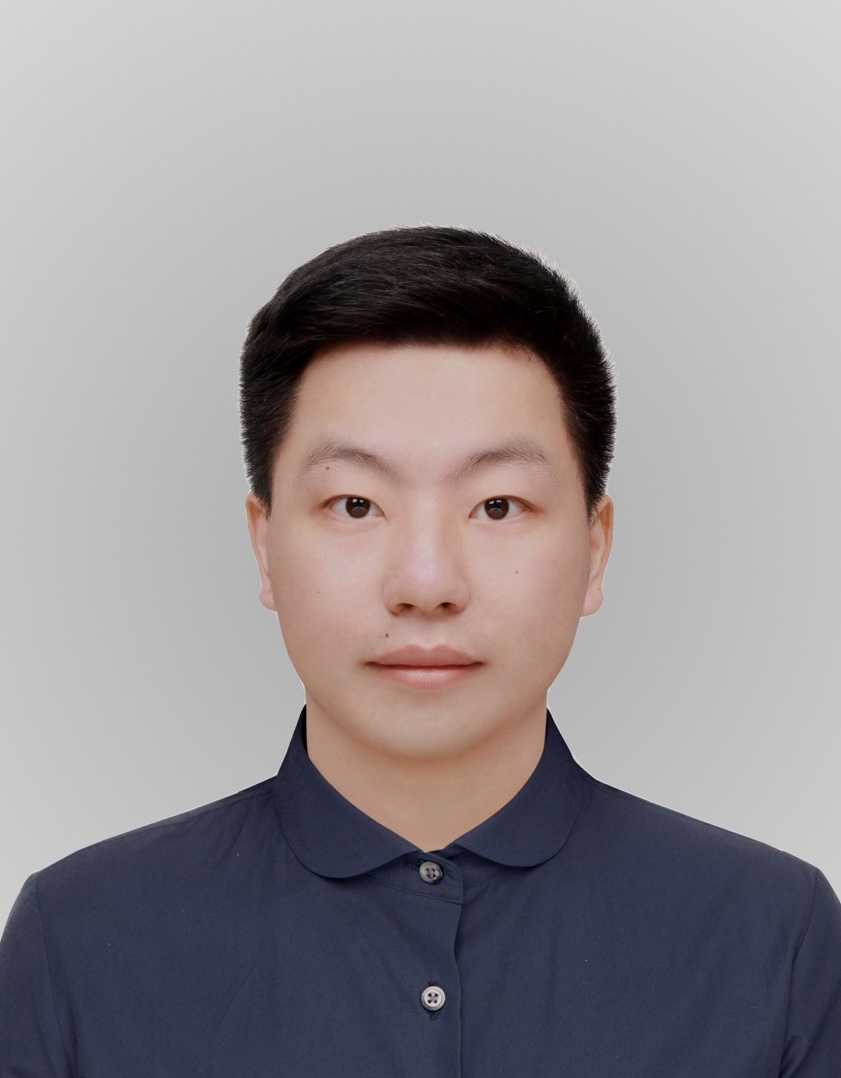}}]
{Sihao Liu} received his B.S. degree in Mathematics and Applied Mathematics (2022) and his M.S. degree in Mathematics (2025) from Guangdong University of Technology, Guangzhou, China. Sihao is currently a Ph.D. student in Computer Science at NC State University. His research interests include networking, Graph Neural Networks, and large-scale optimization.
\end{IEEEbiography}
\begin{IEEEbiography}
[{\includegraphics[width=1in,height=1.25in,clip,keepaspectratio]{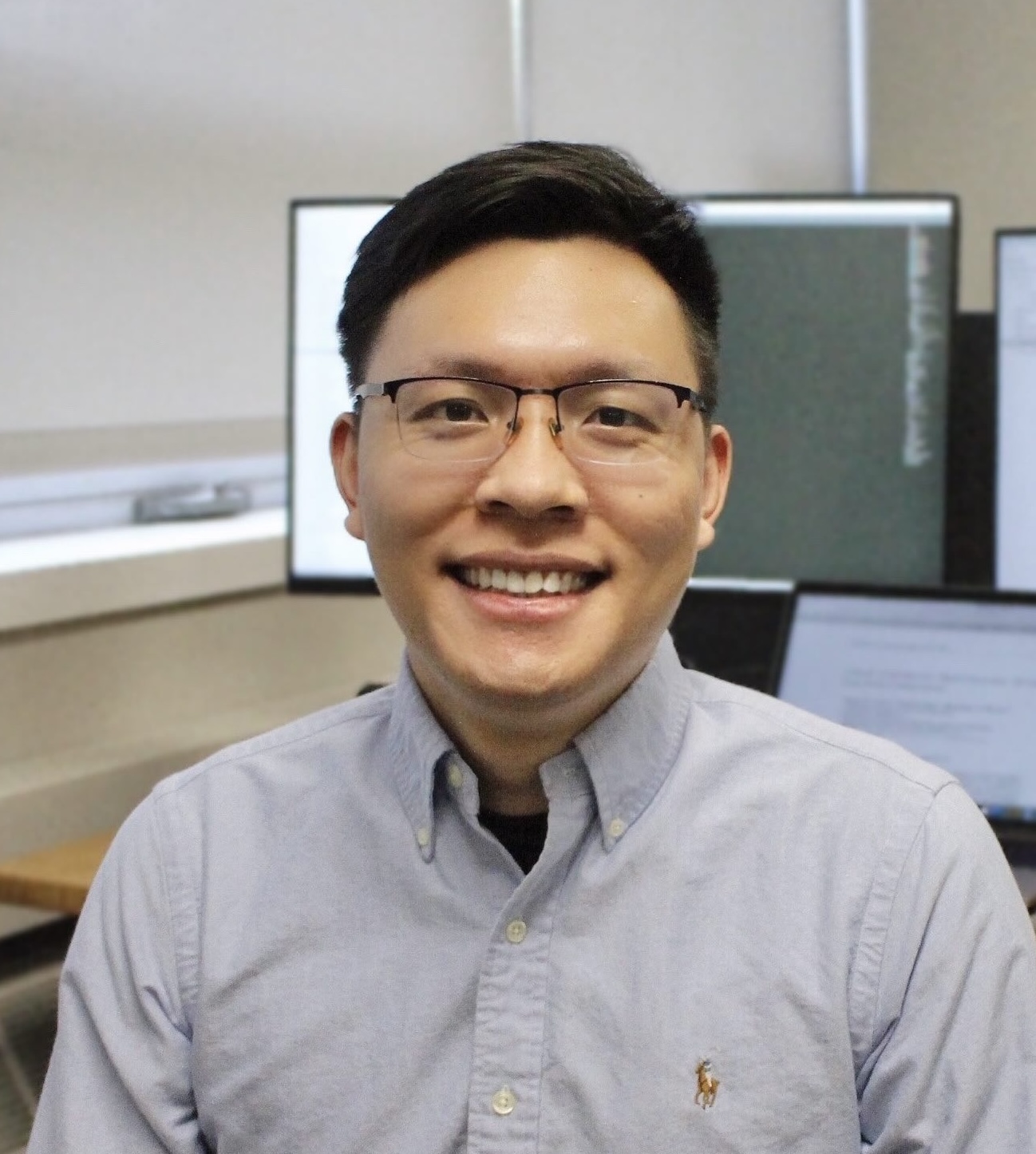}}]
{Xiaorui Liu} (Member 2023) is an Assistant Professor of Computer Science at North Carolina State University (USA). He earned his PhD degree in Computer Science from Michigan State University (USA) in 2022. His research focuses on graph deep learning, trustworthy artificial intelligence, and large-scale machine learning. He is the recipient of NSF CAREER Award (2025), AAAI New Faculty Highlight (2025), ACM SIGKDD Outstanding Dissertation Award (Runner-up, 2023), Amazon Research Award (2023), and Chinese Government Award for Outstanding Students Abroad (2022). He has served on the Organizing Committee of 2025 ACM SIGKDD as a Workshop Chair and as a regular Program Committee member of conferences including ICML, ICLR, NeurIPS, and KDD.
\end{IEEEbiography}
\begin{IEEEbiography}
[{\includegraphics[width=1in,height=1.25in,clip,keepaspectratio]{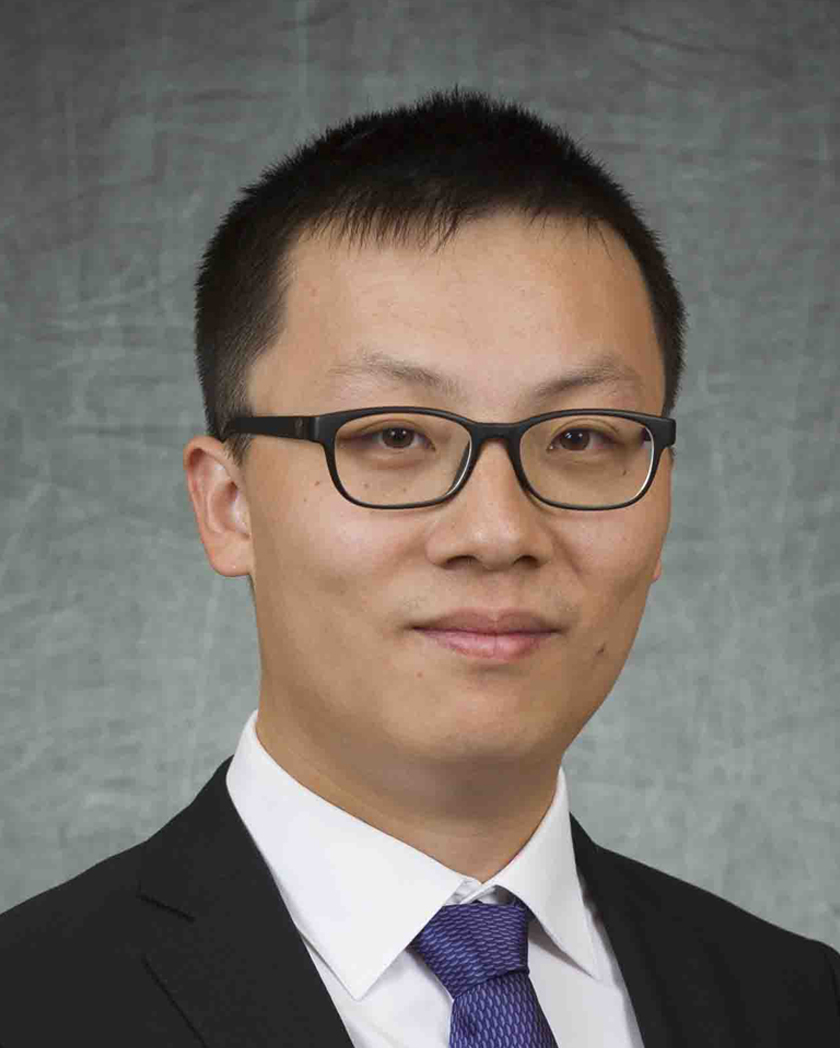}}]
{Ruozhou Yu} (Student Member 2013, Member 2019, Senior Member 2021) is an Assistant Professor of Computer Science at NC State University, USA. He received his PhD degree (2019) in Computer Science from Arizona State University, USA. His interests include quantum networking, edge computing, algorithms and optimization, distributed learning, and security and privacy. He served on the organizing committees of IEEE INFOCOM 2022-2025 and IEEE IPCCC 2020-2024, as a TPC Track Chair for IEEE ICCCN 2023, and as TPC members of IEEE INFOCOM 2020-2026, ACM Mobihoc 2023-2025 and IEEE ICDCS 2025. He served as an Associate Editor for IEEE Transactions on Network Science and Engineering from 2024-2025, and is serving as an Area Editor for Elsevier Computer Networks since 2023. He received the NSF CAREER Award in 2021.
\end{IEEEbiography}
\begin{IEEEbiography}[{\includegraphics[width=1in,height=1.25in,clip,keepaspectratio]{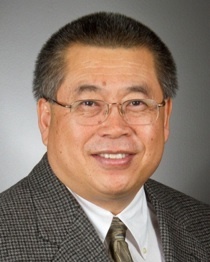}}]
{Guoliang Xue} (Member 1996, Senior Member 1999, Fellow 2011) is
a Professor of Computer Science in the School of Computing and Augmented
Intelligence at Arizona State University.
His research interests span the areas of
Internet-of-things,
cloud/edge/quantum computing and networking,
crowdsourcing and truth discovery,
QoS provisioning and network optimization,
security and privacy,
optimization and machine learning.
He received the IEEE Communications Society William R. Bennett Prize in 2019.
He is an Associate Editor of IEEE Transactions on Mobile Computing,
as well as a member of the Steering Committee of this journal.
He served on the editorial boards of
IEEE/ACM Transactions on Networking
and
IEEE Network Magazine,
as well as the Area Editor of
IEEE Transactions on Wireless Communications, overseeing 13 editors
in the Wireless Networking area.
He has served as VP-Conferences of the IEEE Communications Society.
He is the Steering Committee Chair of IEEE INFOCOM.
\end{IEEEbiography}

\end{document}